# Two years of ALMA bibliography – lessons learned

Silvia Meakins*[a], Uta Grothkopf[a], Marsha J. Bishop[b], Felix Stoehr[a], Ken Tatematsu[c]

[a]European Southern Observatory (ESO), Karl-Schwarzschild-Str. 2, 85748 Garching near Munich, Germany; [b]National Radio Astronomy Observatory (NRAO), 520 Edgemont Road, Charlottesville, VA 22903, USA; [c]National Astronomical Observatory of Japan (NAOJ), 2-21-1 Osawa, Mitaka, Tokyo 181-8588, Japan

## ABSTRACT

Telescope bibliographies are integral parts of observing facilities. They are used to associate the published literature with archived observational data, to measure an observatory's scientific output through publication and citation statistics, and to define guidelines for future observing strategies.

The ESO and NRAO librarians as well as NAOJ jointly maintain the ALMA (Atacama Large Millimeter/submillimeter Array) bibliography, a database of refereed papers that use ALMA data.

In this paper, we illustrate how relevant articles are identified, which procedures are used to tag entries in the database and link them to the correct observations, and how results are communicated to ALMA stakeholders and the wider community. Efforts made to streamline the process will be explained and evaluated, and a first analysis of ALMA papers published after two years of observations will be given.

## 1. INTRODUCTION

During the past years, telescope bibliographies have come to play an indispensable role in the data lifecycle of an observatory. Based on the observational data obtained by the researchers, authors publish their results in a variety of journals. Telescope bibliographies are compilations of such papers. Typically, telescope bibliography databases provide a public user interface and allow the bibliography compilers as well as interested users to derive reports and statistics about the content.

The ALMA (Atacama Large Millimeter/submillimeter Array) bibliography is a database of refereed papers published by the ALMA users community that use partly or exclusively ALMA data. Since ALMA is a partnership between North America, Europe, and East Asia in cooperation with the Republic of Chile, also the bibliography is a joint project of the ESO and NRAO librarians, together with NAOJ. At ESO and NRAO, considerable experience has been accumulated in the past with regard to the maintenance and further development of telescope bibliographies thanks to the bibliographies at the respective observatories: NRAOpapers[1] and ESO telbib[2]. This existing knowledge, along with the infrastructure available at both institutions, was of primary importance in establishing the ALMA bibliography.

The fact that librarians, astronomers, and archive specialists have cooperated intensely from early project stages on by exchanging ideas and experiences was crucial in order to be ready to track and process papers as soon as the first publications appeared. This joint approach underlines the general spirit of collaboration that also characterizes the ALMA project at large.

*smeakins@eso.org; phone +49 89 32006-775; http://www.eso.org/libraries/

---

[1] http://library.nrao.edu/papersmethod.shtml
[2] http://www.eso.org/sci/libraries/telbib_info.html

## 2. ALMA BIBLIOGRAPHY WORKFLOW

Maintaining the ALMA bibliography involves various steps that are necessary in order to assure that the database is as complete and accurate as possible.

### 2.1 Access to data by PI

Once data have been taken, they are made available to the PI. A proprietary period of 12 months is applied, starting at the time when data is delivered to the PI. An exception to this rule are observations made during Director's Discretionary Time project, for which the period is six months. During this time, the PI is the only authorized user. Access is provided through the ALMA archive.

### 2.2 Using ALMA data for publications

ALMA has issued a clear policy in case observations are used in publications. This policy is communicated to researchers repeatedly and in various ways, e.g. in communications with PIs when they get access to the data, on the ALMA data webpage[3] and Knowledgebase/FAQ[4], and in case data are retrieved from the archive.

The official data-citing statement that has to be included in the acknowledgement section of each paper is as follows:

"*This paper makes use of the following ALMA data: ADS/JAO.ALMA#2011.0.01234.S. ALMA is a partnership of ESO (representing its member states), NSF (USA) and NINS (Japan), together with NRC (Canada) and NSC and ASIAA (Taiwan), in cooperation with the Republic of Chile. The Joint ALMA Observatory is operated by ESO, AUI/NRAO and NAOJ.*"

The project code 2011.0.01234.S is a placeholder that needs to be replaced with the code of the actual observations.

In addition, publications from North American authors must include the standard NRAO acknowledgement:

"*The National Radio Astronomy Observatory is a facility of the National Science Foundation operated under cooperative agreement by Associated Universities, Inc.*"

According to our experience so far, the ALMA users community has accepted the citation policy very willingly and widely. The high fraction of papers that clearly mention ALMA project codes of the observations used in the publications helps the bibliography curators enormously in detecting and correctly tagging ALMA papers.

### 2.3 Compiling the ALMA bibliography

The bibliography is compiled by scanning the major astronomical journals for papers that use ALMA data. The curators make extensive efforts in order to identify all relevant papers, including screening of manuscripts posted on the astro-ph eprint server[5] and use of the semi-automated full-text search tool FUSE (developed at ESO) which scans defined sets of journal articles for specific keywords and project codes and provides highlighted results in context (Fig. 1).

Journals currently included in the regular screenings are *A&A, A&ARv, AJ, AN, ApJ, ApJS, ARA&A, EM&P, Icarus, MNRAS, Nature, NewA, NewAR, PASJ, PASP, P&SS,* and *Science*. In addition, several other journals as well as preprints on astro-ph are checked. This way, each year more than 10,000 papers are scanned.

---

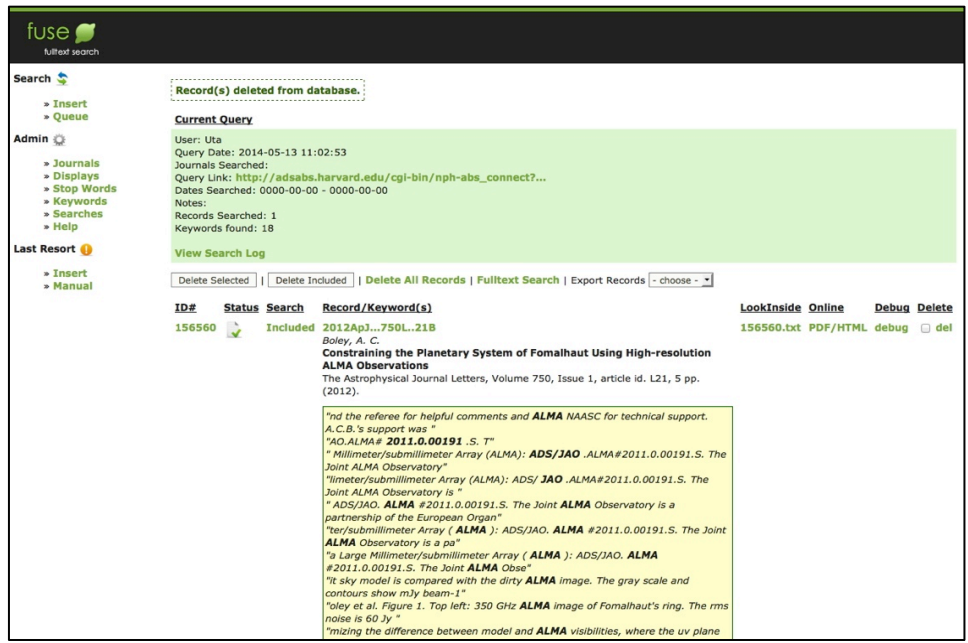

Figure 1. The FUSE full-text search tool is used to screen journals for ALMA-related keywords and project codes.

## 2.4 Policy for paper inclusion and data verification

All papers for which at least one ALMA-related keyword was retrieved are then visually inspected by the librarians to decide whether or not they shall be included in the database. In order to be eligible for the ALMA bibliography, papers have to fulfill certain criteria:

- ALMA data must have been used either alone or in combination with other data; these can be proprietary or archival data

- Papers are excluded if they

  - merely quote results from the literature

  - describe instrumentation or software

  - simply mention ongoing projects

  - suggest future observations

  - develop models or run simulations without use of data

Bibliographic metadata such as author names and affiliations, titles, and publication details are imported into the bibliography from the NASA Astrophysics Data System (ADS)[6]. All project codes provided by the authors are crosschecked with the ALMA archive in order to make sure that no typos or other mistakes occur. In case of inconsistencies, various steps are taken to identify the correct codes, including consultation with ALMA scientists and communication with the authors.

## 2.5 ALMA bibliography parameters

Once a project code has been identified, it is entered by the librarians through the ALMA bibliography backend. Further information is retrieved on-the-fly from the ALMA science archive: (i) the ALMA partner during whose observing time the data were taken, and (ii) the observing type, such as Large, Target of Opportunity, Director's Discretionary Time, and Science Verification. Project codes of science verification observations are assigned to the four partners Joint ALMA Office (JAO), Europe, North America, and East Asia.

For ALMA papers, it has been agreed to track merely the use of the telescope at large rather than the individual receiver bands. This is different from ESO's and NRAO's existing bibliographies of their facilities, where individual instruments are traced and recorded.

An important parameter added to bibliography records by the librarians are tags indicating whether data were obtained by the team of authors or retrieved from the archive. If there is no overlap between authors and PI/CoIs, a paper is regarded as making use of archival data. An exception are papers with data from science verification programs which are always tagged archival due to the fact that they are made available to the entire community immediately upon their ingestion into the archive.

In this context, a close collaboration between bibliography curators and archive scientists is particularly important in order to achieve an efficient workflow. For instance, a feature has been implemented in the ALMA bibliography backend that allows the librarians to automatically cross-check PI/CoI names (obtained from the proposal information in the archive) with all names of authors of the paper. In case no overlap is noticed, the system suggests tagging this paper as archival. This suggestion can be accepted or rejected; the final decision remains with the curators.

## 2.6 Public interface

A public interface will be made available via the ALMA Science Portal in the near future. At present, both ESO and NRAO provide access to the ALMA bibliography through their respective databases[7,8].

Along with bibliographic details (authors, title, journal, etc.), entries in the bibliography show the ALMA data project codes as well as the respective ALMA partners. Project codes are linked to the ALMA archive, so that interested users can easily move over to the Science Archive query form to submit a download request for the corresponding data. In future, data retrieved through archive queries will link back to the ALMA bibliography so that all papers that have been published so far with the respective observations can be seen.

A similar workflow at ESO, using the search tool FUSE and the ESO telescope bibliography (telbib), has for instance, been described in [1] and [2].

# 3. REPORTS AND STATISTICS

In addition to the public interface, statistics derived from the ALMA bibliography are communicated to the astronomy community as well as to ALMA stakeholders in a number of other ways, most notably through inclusion of the results in print and online reports such as the *Basic ESO Publication Statistics*[9] and the *NRAO Publication Statistics*[10]. In the following, we present selected results. The librarians at NRAO and ESO are available to answer more detailed questions.

## 3.1 Number of papers and journal coverage

By the end of 2013, a total of 85 refereed papers were published using ALMA data (2012: 20; 2013: 65). ALMA publications typically experience a delay of less than a year (including analysis, writing the paper, the refereeing process and the publication itself). Despite the fact that not all PIs received the data, the fraction of published data is quite high already (around 70%). The largest fraction of papers appeared in *ApJ* (almost 52%), followed by *A&A* (approx. 24%) and *MNRAS* (more than 7%). A remarkably high number of publications (8%) have been published in the high-impact journals *Nature* and *Science* (Fig. 2).

## 3.2 Science categories and proposal types

When submitting proposals, authors have to indicate the scientific category to which the proposed observations belong. In Fig. 3, the fraction of scientific categories of the data used in papers is shown. Currently about half of the publications are done with data from the "ISM-Star formation" category, which was also the category that received the largest request for time.

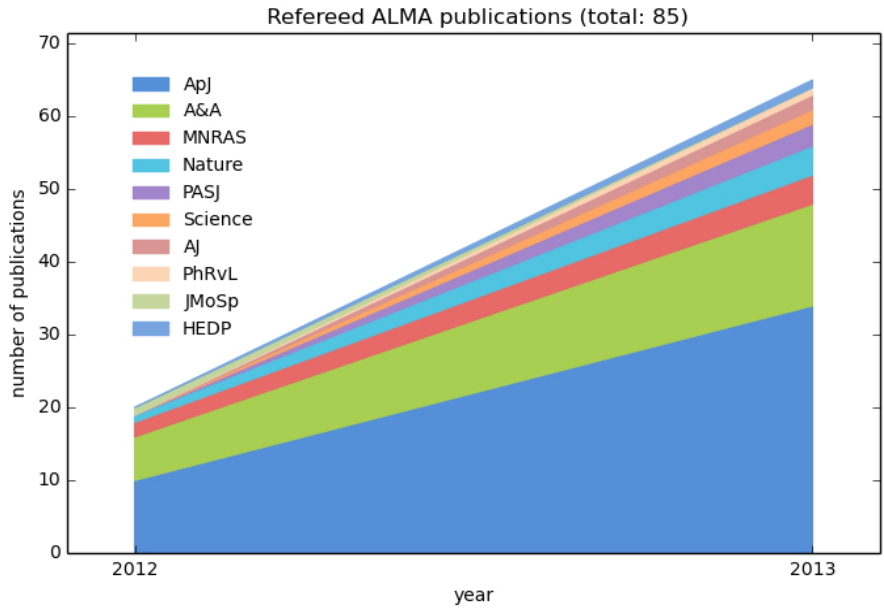

Figure 2. Number of papers per journal published in 2012 and 2013

Note that the definition of the scientific categories has changed between Cycle 0 and Cycle 1+. In Cycle 0, the categories were 1. Cosmology and the high redshift universe, 2. Galaxies and galactic nuclei, 3. ISM, star formation/protoplanetary disks and their astrochemistry, exoplanets, 4. Stellar evolution, the Sun and the solar system. In Cycle 1+, categories 1 and 2 stayed the same, but 3 and 4 were redefined (because of the number of proposals received in those panels) as follows: 3. ISM, star formation and astrochemistry, 4. Circumstellar disks, exoplanets and the solar system, 5. Stellar evolution and the Sun. As it is impossible for us to find out for each Cycle 0 program to which Cycle 1+ category it would belong, we treat the Cycle 0 categories as if they were the Cycle 1+ categories.

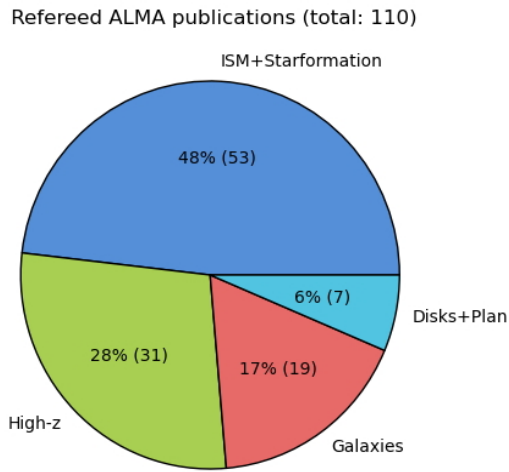

Figure 3. Number of papers by scientific category of the data they used. Publication years: 2012 – Apr 2014.
See text for an explanation of the categories.

### 3.3 Re-publication of ALMA data

Already, there are various projects that have been published several times. The largest numbers of republications stem from one science verification and one standard project which have produced 9 papers each by April 2014. In Fig. 5, the number of published ALMA projects is shown as a function of the number of publications in which they appear.

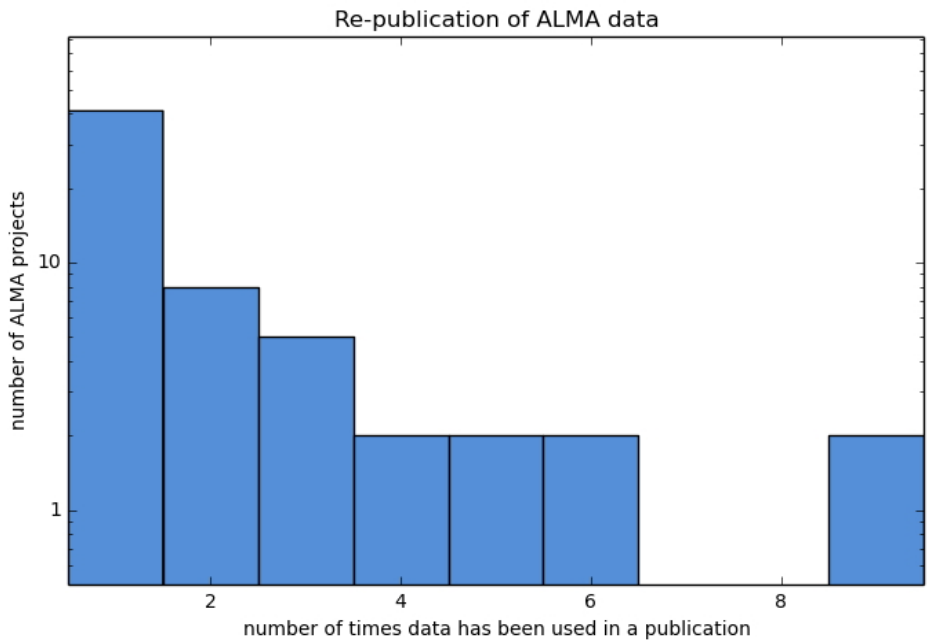

Figure 5. Number of published ALMA projects as a function of the number of publications they appear in.
Total=110, publication years: 2012 – Apr 2014. Note the log y-axis.

## 4. LESSONS LEARNED

After two years of collaborative maintenance of the ALMA bibliography, we can draw first conclusions regarding the methodology, workflow, and policies we applied.

### 4.1 Regular communication among curators

Any project with contributors located on different continents requires frequent communication among the partners involved. This is particularly true for the compilation of a database, where it is essential to agree on specific policies and procedures to which all groups adhere. For the ALMA bibliography, we have established regular exchange regarding the content of the database among the core curators, i.e., the librarians of NRAO and ESO as well as NAOJ. In addition, ALMA archive specialists and ALMA Management are consulted whenever questions arise or general policies need to be defined or clarified. According to our experience, frequent communication among all groups is essential to create and maintain a bibliography that is as complete, consistent, and yet multi-faceted as possible.

### 4.2 Using tools to streamline the process

During the next years, the ALMA bibliography curators will be confronted with an ever-increasing number of papers that need to be reviewed and ingested into the database. It is essential to streamline processes whenever possible. The use of tools such as the semi-automatic full-text search tool FUSE are not only saving time, but are also indispensible in order to standardize procedures and make the workflow more efficient. Moreover, they facilitate the identification of relevant papers enormously by being able to detect occurrences of keywords at almost zero-error rate, a precision human inspectors cannot possibly achieve.

### 4.3 Communication with PIs/CoIs

In order to raise awareness among the ALMA users community about policies that have been established, bibliography curators have to communicate clearly and repeatedly to PIs and CoIs how data should be acknowledged in papers. It is highly advisable to provide authors with precise wording that can be copied and pasted into the acknowledgements section of papers.

Early inspection of papers, i.e., before they are actually published, can help greatly in this process. In case authors forgot to give proper credit of data provided by the observatory, a review of manuscripts (eprints) and subsequent communication with authors can lead to a corrected version being published rather than the original one with incomplete or no acknowledgements. This procedure also has a certain "educational" effect, as authors will typically be even more careful during the preparation of future papers.

### 4.4 Visual inspection of papers

Visual inspection of papers by the bibliography curators is an absolute necessity. Software alone, no matter how sophisticated, will not be able to understand all subtleties when authors describe if and how they used observations. Hence, any bibliography that is exclusively machine-generated will both miss papers (if authors by mistake omit information necessary in order to correctly identify observations) and will also contain wrong hits, i.e., papers that mention the correct keywords, but in an irrelevant context.

### 4.5 Key performance indicators: implement early on!

Telescope bibliographies are key-performance indicators as they help greatly to understand the scientific productivity (as expressed in number of publications) and the impact (typically derived from the number of citations) of a facility. As such, they form an integral part of the overall workflow and have to be designed, developed and implemented from the very beginning of an observatory's life cycle. Delayed creation inevitably leads to loss of time as re-engineering of processes will be necessary, bears a high risk of incompleteness of the bibliography, and ultimately results in higher costs and diminished efficiency.

## ACKNOWLEDGEMENTS

We are very grateful to Chris Erdmann, now at the Harvard-Smithsonian Center for Astrophysics, who created the first versions of the ESO Telescope Bibliography (telbib)/NRAOpapers and the FUSE full-text search tool. The ALMA bibliography and FUSE use NASA's Astrophysics Data System (ADS) Abstract Services; the authors wish to thank the ADS team for their excellent service to the astronomy community.